\documentclass[useAMS,usenatbib]{mn2e}
\usepackage{amsmath}
\usepackage{amssymb}

\usepackage{graphics}
\usepackage{graphicx}
\usepackage{epstopdf}
\usepackage{footnote}
\usepackage{tablefootnote}
\usepackage{txfonts}
\usepackage{lipsum}
\usepackage{float}
\usepackage[draft]{hyperref}
\usepackage{macros}
\usepackage{fixltx2e}

\usepackage{titlesec}

\titlespacing*{\section}
{0pt}{1ex }{1ex}
\titlespacing*{\subsection}
{0pt}{1ex }{1ex}
\vspace{-0.3cm}
\makesavenoteenv{tabular}

\DeclareSymbolFont{cmletters}{OML}{cmm}{m}{it}
\DeclareMathSymbol{v}{\mathalpha}{cmletters}{"76}

\usepackage[usenames,dvipsnames,svgnames,table]{xcolor}
\usepackage{hyperref}
\definecolor{darkblue}{rgb}{0.0,0.0,0.3}
\hypersetup{colorlinks,breaklinks,
            linkcolor=darkblue,urlcolor=darkblue,
            anchorcolor=darkblue,citecolor=darkblue}

\title[MHD Wind Accretion Onto Sgr A*]{Accretion of Magnetized Stellar Winds in the Galactic Centre: Implications for Sgr A* and PSR J1745-2900 }
\author[S. M. Ressler, E. Quataert, J. M. Stone ]{S. M. Ressler$^{1},$  E. Quataert$^{1},$ J. M. Stone$^{2}$\\
$^{1}$Departments of Astronomy \& Physics, Theoretical Astrophysics Center, University of California, Berkeley, CA 94720 \\
$^{2}$Department of Astrophysical Sciences, Princeton University, Princeton, NJ 08544}

\begin{document}

\maketitle

\begin{abstract}
The observed rotation measures (RMs) towards the galactic centre magnetar and towards Sagittarius A* provide a strong constraint on MHD models of the galactic centre accretion flow, probing distances from the black hole separated by many orders of magnitude.  We show, using 3D simulations of accretion via magnetized stellar winds of the Wolf-Rayet stars orbiting the black hole, that the large, time-variable RM observed for the pulsar PSR J1745-2900 can be explained by magnetized wind-wind shocks of nearby stars in the clockwise stellar disc.  
In the same simulation, both the total X-ray luminosity integrated over 2-10$''$, the time variability of the magnetar's dispersion measure, and the RM towards Sagittarius A* 
are consistent with observations.   We argue that (in order for the large RM of the pulsar to not be a priori unlikely) the pulsar should be on an orbit that keeps it near the clockwise disc of stars.  We present a 2D RM map of the central 1/2 parsec of the galactic centre that can be used to test our models. Our simulations predict that Sgr A* is typically accreting a significantly ordered magnetic field that ultimately could result in a strongly magnetized flow with flux threading the horizon at $\sim$ 10$\%$ of the magnetically arrested limit.
\end{abstract}

\begin{keywords}
Galaxy: centre -- accretion, accretion discs -- (magnetohydrodynamics) MHD  -- stars: Wolf-Rayet -- polarization -- black hole physics
\end{keywords}

\vspace{-1cm}

\section{INTRODUCTION}

The two largest rotation measures (RMs) observed in the galaxy are located within the central $\sim$ 0.1 pc of Sagittarius A* (Sgr A*).  The largest is towards the radio source Sagittarius A* associated with the $\sim 4.3 \times 10^6$ $M_\odot$ black hole, which was measured in 2005 over a two-month time frame to be $\approx -5.6 \times 10^5$ rad/m$^2$ \citep{Marrone2007}. This value was roughly constant in time over that short interval and has had the same sign for at least $\sim$ 5 years before then \citep{Bower2002}. 
The second largest RM is that observed towards the magnetar PSR J1745-2900 \citep{Eatough2013} with a value of $\approx -6.6 \times 10^4$ rad/m$^2$.
If, as is generally supposed, these large RMs are produced locally to the galactic centre, they offer the most direct probes of the magnetization of the accretion flow, a critical parameter for determining the state of the accreting plasma at Schwarzschild radii scales.

Both analytic modeling \citep{Quataert2004,Shcherbakov2010} and three dimensional hydrodynamic simulations \citep{Cuadra2008,Russell2017,Ressler2018}, have shown that the winds of the $\sim$ 30 Wolf-Rayet (WR) stars orbiting Sgr A* can account for the amount of hot gas observed by \emph{Chandra} in X-rays at distances $\lesssim 10''$ from the central black hole \citep{Baganoff2003}.  This hot gas has been invoked to explain the large RM observed in PSR J1745-2900 and even used as evidence that the galactic centre accretes strongly magnetized plasma \citep{Eatough2013}.  If that is indeed the case, the winds of the stars themselves are the most likely source of magnetic field; any ambient field that may have been present would have been blown away by the winds.
Unfortunately, though the mass-loss rates and wind speeds of the stars are reasonably well constrained observationally \citep{Martins2007,YZ2015}, nothing is known about the structure or the magnitude of the magnetic field in the winds.  However, given that the orbital velocities and 2D positions of the stars are also known \citep{Paumard2006,Lu2009,Gillessen2017}, the problem of explaining the RM of Sgr A* and the magnetar is still fairly well posed.  

In this letter, we present the first three dimensional, magneto-hydrodynamic (MHD) simulations of the accretion flow around Sgr A* that include the winds of the WR stars. Our model extends that of \citet{Ressler2018} (hereafter R18) to include magnetized stellar winds, allowing for self-consistent modeling of the X-ray emission, the RMs towards PSR J1745-2900 and Sgr A*, and the inner accretion flow onto Sgr A*; we defer a detailed study of the latter
to a future paper. 
\S \ref{sec:wind_model} describes the physical model for the stellar winds employed in our simulation, \S \ref{sec:analytic} describes a simple toy model of isolated stellar winds useful for interpreting the RM towards the pulsar, \S \ref{sec:MHD} presents the results of the full 3D simulation of magnetized stellar wind accretion onto Sgr A*, focusing on the two RMs, and \S \ref{sec:conc} concludes. 

\section{MAGNETIZED WIND MODEL}
\label{sec:wind_model}
The hydrodynamic stellar wind model described in R18 treats stellar winds as source terms in mass, momentum, and energy that move on fixed Keplerian orbits through the simulation domain.  The sources are $r_w \approx$ $2\sqrt{3} 
\Delta x$ in radius, where $\Delta x$ is the local grid spacing evaluated at the centre of the ``star.'' This model was shown to accurately drive a wind possessing the desired mass loss rate, $\dot M_w$, and constant radial velocity, $v_{w}$, with negligible temperature.  In order to make these winds magnetized, in this work we add two additional source terms to our {\tt Athena++} simulations: one in the induction equation and one in the total energy equation.

We expect the magnetic fields of stellar winds at distances $r'$ $\gg$ the Alfv\'en radius, $r_A$, the point at which the magnetic energy density of the wind equals its kinetic energy density, to be predominately in the $\varphi'$ direction, where primes denote the frame of the star with the $z'$-axis aligned with its rotation axis. This is because flux conservation requires that the radial component of the field, $B_{r'}$, scales as $(r')^{-2}$, while the azimuthal component, $B_{\varphi'}$, scales as $(r')^{-1}$, so that at large radii $B_{\varphi'}$ will ultimately be the dominant component of the field \citep{WeberDavis}.

Here we parameterize the field by the ratio between the wind's ram pressure and magnetic pressure evaluated in the equatorial plane at $r'=r_A$, $\beta_w  \equiv \left. 8 \pi \rho v_w^2/B_{\varphi'}^2 \right|_{\theta' = \pi/2}= 2 \dot M v_w/(B_A^2 r_A^2) = \textrm{const.}  $, where $B_A$ is the magnitude of the magnetic field at $r' = r_A$ and $\theta' = \pi/2$. For $r_A = R_\odot$, $M_w = 10^{-5}$ $M_\odot$/yr, and $v_w = 1000$ km/s, a given $\beta_w$ corresponds to $B_A\approx 5.1$ kG/$\beta_w^{1/2}$. Though observational estimates of the magnetic fields in the winds of WR stars are sparse, $\sim$ 10$\%$ of O-stars have been observed to have surface fields as high as $\sim 100$G-$20$kG (e.g. \citealt{Donati2009,Wade2016}), so we expect this value of $B_A$ to be reasonable for at least some of the galactic centre stars.

Adding a source term to the induction equation while maintaining $\mathbf{\nabla}\cdot \mathbf{B}=0$ 
requires precise consistency with the constrained transport algorithm used by {\tt Athena++} to avoid numerical instability. Therefore, instead of adding a source term directly to the magnetic field, for each star we instead add a source term, $\mathbf{E_w}$, to the electric field with a curl only in the $\varphi'$-direction:
\begin{equation}
\mathbf{E_w} = -\frac{\pi B_A r_A v_w}{  r_w^2} \cos(\theta) \sin\left(\frac{r'}{r_w} \pi\right) \mathbf{r'} 
\label{eq:Ew}
\end{equation}   
for $r'<r_w$ and 0 otherwise.
This electric field  corresponds to a source of magnetic field for $r'<r_w$ 
\begin{equation}
\mathbf{\nabla} \times \mathbf{E_w} = \frac{\pi B_A r_A v_w}{  r_w^2} \sin(\theta) \sin\left(\frac{r'}{r_w} \pi\right) \hat\varphi',
\end{equation}
and 0 otherwise.
The radial dependence of the electric field in Equation \eqref{eq:Ew} was chosen to ensure that the field is continuous at the boundary of the source at $r'=r_w$, while the angular dependence was chosen to ensure that hoop stress doesn't diverge at the poles.
This source of magnetic field also sources the total energy equation 
\begin{equation}
\dot E_B = \frac{1}{4\pi} B_{\varphi'} \left(\mathbf{\nabla} \times \mathbf{E_w}\right) \cdot \hat \varphi',  
\end{equation}
again for $r'<r_w$ and 0 otherwise.  In each cell, $\dot E_B$ is volume-averaged while $\mathbf{E_w}$ is averaged over the appropriate cell edge (see Equations 22-24 of \citealt{Stone2008}). These source terms, in addition to the point source gravity of the black hole, optically thin radiative cooling due to line and bremsstrahlung emission, and the hydrodynamic wind source terms described in more detail in R18 are added to the conservative MHD equations.

For $\beta_w \gtrsim 5$, this model successfully drives a wind with the desired $\dot M_w$, $v_w$, and $\beta_w$ while retaining the $\sin(\theta)$ dependence of the magnetic field.  For $\beta_w \lesssim5$, however, magnetic pressure serves to accelerate the wind in the radial direction.  Thus, decreasing the parameter $\beta_w$ beyond $\sim$ 5 does not ultimately end up increasing $\left. B_\varphi^2/(8 \pi \rho v^2) \right|_{\theta'=\pi/2}$, which saturates at $\sim 0.2$.  This is not just a limitation of our simple model but a physical limitation on the magnetization of winds at large radii (e.g. \citealt{Lamers1999}).   Though a more sophisticated treatment of the angular dependence of $B_{\varphi'}$ might result in a slightly different saturation value, in general we expect $\beta_w \gtrsim 1$.  Therefore, in what follows we consider only $\beta_w \ge 1$ in our analytic calculations and $\beta_w \ge 10$ in our simulations.

\section{ANALYTIC MODEL FOR THE RM OF PSR J1745-2900 } 
Assuming a standard spherically symmetric wind with a toroidal magnetic field parameterized by $\beta_w$ as in \S 2, the RM for a single stellar wind is given by
\begin{equation}
\begin{aligned}
\textrm{RM}_* \approx& \frac{15000 \textrm{ rad m}^{-2}}{\beta_w^{1/2}} \left(\frac{\dot M_w}{10^{-5} M_\odot /\textrm{yr}}\right)^{3/2} \left(\frac{s}{10^{-2} \textrm{ pc}}\right)^{-2} \left(\frac{v_w}{10^3 \textrm{ km/s}}\right)^{-1/2} \\ 
&\times \int\limits_{-\infty}^{z_p} \frac{s^2 \sin(\theta')\hat \varphi ' \cdot \hat z }{(s^2 + z^2)^{3/2}} dz, 
\end{aligned}
\label{eq:RM_cgs}
\end{equation} 
where the positive $z$-direction points away from Earth,  while $z_p$ and $s$ are, respectively,  the $z$-coordinate of and the projected distance to the pulsar. Since the dimensionless integral in Equation \eqref{eq:RM_cgs} can take on any value between $\pm$ $\pi/2$,, we have
\begin{equation}
|\textrm{RM}_*| \lesssim  \frac{23000 \textrm{ rad m}^{-2}}{\beta_w^{1/2}} \left(\frac{\dot M_w}{10^{-5} M_\odot /\textrm{yr}}\right)^{3/2} \left(\frac{s}{10^{-2} \textrm{ pc}}\right)^{-2} \left(\frac{v_w}{10^3 \textrm{ km/s}}\right)^{-1/2}.
\label{eq:RM_max}
\end{equation}

Equation \eqref{eq:RM_cgs} shows that the RM for a given star is a rapidly decreasing function of projected distance, $\textrm{RM}_*$  $\tilde{\propto}$ $s^{-2}$, so that only the stars closest to the line of sight (LOS) will significantly contribute.  Furthermore, it shows that in order for RM$_*$ to be on the order of the observed -6.6 $\times 10^4$ rad/m$^{2}$, there needs to be a star located $\lesssim 10^{-2}$ pc in projected distance from the pulsar assuming values typical of WR stars for $\dot M_w$ and $v_w$. The closest WR star (E32 aka 16SE1), however, has $s \sim 2.5 \times 10^{-2}$ pc and even with optimistic assumptions for other parameters would require a very large mass loss rate, $\dot M_w \approx 7 \times 10^{-5} M_\odot$/yr to reach $|\textrm{RM}|_* \sim 6.6 \times 10^4$ rad/m$^{2}$.

Therefore, we conclude that it is unlikely that isolated stellar winds can produce a RM as large as that observed for the galactic centre magnetar. However, the RM near a star can be enhanced by a factor as much as $\sim 16$ or more by the presence of shocks with other winds or with the ambient medium.  In fact, there are two other stars in the near vicinity of E32, namely, E23 (aka 16SW), and E40 (aka 16SE2),
both located within $\sim 0.01$ pc in projected distance from E32.  Since all three are disc stars, they are also clustered in 3D positions. As we now show, shocks between these stars are then expected and will affect the RM of the pulsar.

\label{sec:analytic}
\section{3D MHD SIMULATIONS}
\label{sec:MHD}
\subsection{Parameter Choices and Computational Grid}
The ``stars'' in our simulation are on fixed Keplerian orbits using the same prescription described in R18,
where the $z$-coordinate of a star in the year $\sim 2005$, $z_*$, is taken from \citet{Paumard2006} 
for stars within the stellar disc, while $z_*$ for a non-disc star is set so as to minimize the eccentricity of its orbit.  We use the mass loss rates and wind speeds of \citet{Cuadra2008} for all of the stars except for E23, E32, and E40,
which we allow to vary within a range of uncertainty while fixing $\beta_w = 10$.  

We ran a suite of simulations with different random choices for the spin axis directions of the stars and the mass loss rates and wind speeds of E23, E32, and E40. 
Here we present only one of these simulations, hereafter referred to as ``the fiducial model;'' out of the 7 random variations in wind properties we tried, this was the simulation that best reproduced the observed RM of the pulsar.  We emphasize that this model is not unique and that our purpose is not to do a full parameter survey but to show that a reasonable choice of wind parameters can indeed reproduce the observed pulsar RM.  Furthermore, some of our results depend on the choice of $\beta_w$, with the RM of the pulsar roughly scaling as $\sim$ $\beta_w^{-1/2}$, while RM of Sgr A* and the net flux threading the inner boundary are only weakly dependent on $\beta_w$, perhaps because of magnetic field amplification at small radii.  In a future paper we will explore other models.

The parameters of the three stellar winds closest to the magnetar (which sets its RM in our calculations) for this fiducial model are shown in Table \ref{table:stars}, where we have denoted the spin axes of the stars (which determine the direction of the magnetic fields in the winds) as $\mathbf{n} = (n_x,n_y,n_z)$, defined in the same coordinate system as \citet{Paumard2006}.
Though the spin axes for the remaining stars are just as important for determining the RM of Sgr A*, there is not as direct a relationship between their values and the resulting RM compared to the case of the pulsar.
The values of $\dot M_w$ and $v_w$ shown in Table \ref{table:stars} are all within reasonable systematic observational uncertainties and do not significantly alter the total X-ray luminosity found in R18 that agrees well with \emph{Chandra} observations, nor do they add any local X-ray excess that would have previously been observed near the pulsar.  

Our computational grid is a 1 pc$^3$ box in Cartesian coordinates, with a base resolution of 128$^3$ in addition to 8 levels of nested static mesh refinement, emulating a grid with  logarithmically spacing in radius.  A region with radius of $\approx$ twice the smallest grid spacing at the centre of the grid, $r_{in} \approx $ 1.2 $\times 10^{-4}$ pc is set to floors in density and pressure with zero velocity.  The magnetic field is allowed to freely evolve in this region.  In addition to the the floors and ceilings on density, temperature, pressure, and velocity described in R18, we add a density floor such that $B^2/(4\pi \rho) \le \sqrt{2GM_{BH}/r_{in}}$, where $M_{BH} \approx 4.3 \times 10^6 M_\odot$ is the mass of Sgr A*.  This condition is only activated at the innermost radii in magnetically dominated polar regions.   As in R18, we run the simulation for $1.1$ kyr up to what we refer to as the present day, $t=0$, defined as January 1, 2017.

\begin{table}
\caption{Mass Loss Rates, Winds Speeds, and Spin Axes of the Three Stars Closest to PSR J1745-2900}
\begin{tabular} {|r|r|r|r|r|r|r|}
\hline
Name&Alt. Name &$\dot M_{w}$& $v_{w}$&$n_x$ & $n_y$& $n_z$\\
\hline
E23&16SW&      0.8$\times 10^{-5}$& 440&0.06 & -0.70&0.71\\
E32&16SE1&       2.7$\times 10^{-5}$& 435& -0.08& -0.88& 0.47\\
E40&16SE2&       6.3$\times 10^{-5}$& 1220&-0.22&0.95& -0.23 \\
\hline
\end{tabular}
\label{table:stars}
Notes-- $\dot M_w$ is measured in $M_\odot$/yr and $v_w$ is measured in km/s.  Names are from \citet{Paumard2006}.  
 \\
\end{table}

\subsection{Rotation Measure of PSR J1745-2900}

\begin{figure}
\includegraphics[width=0.4\textwidth]{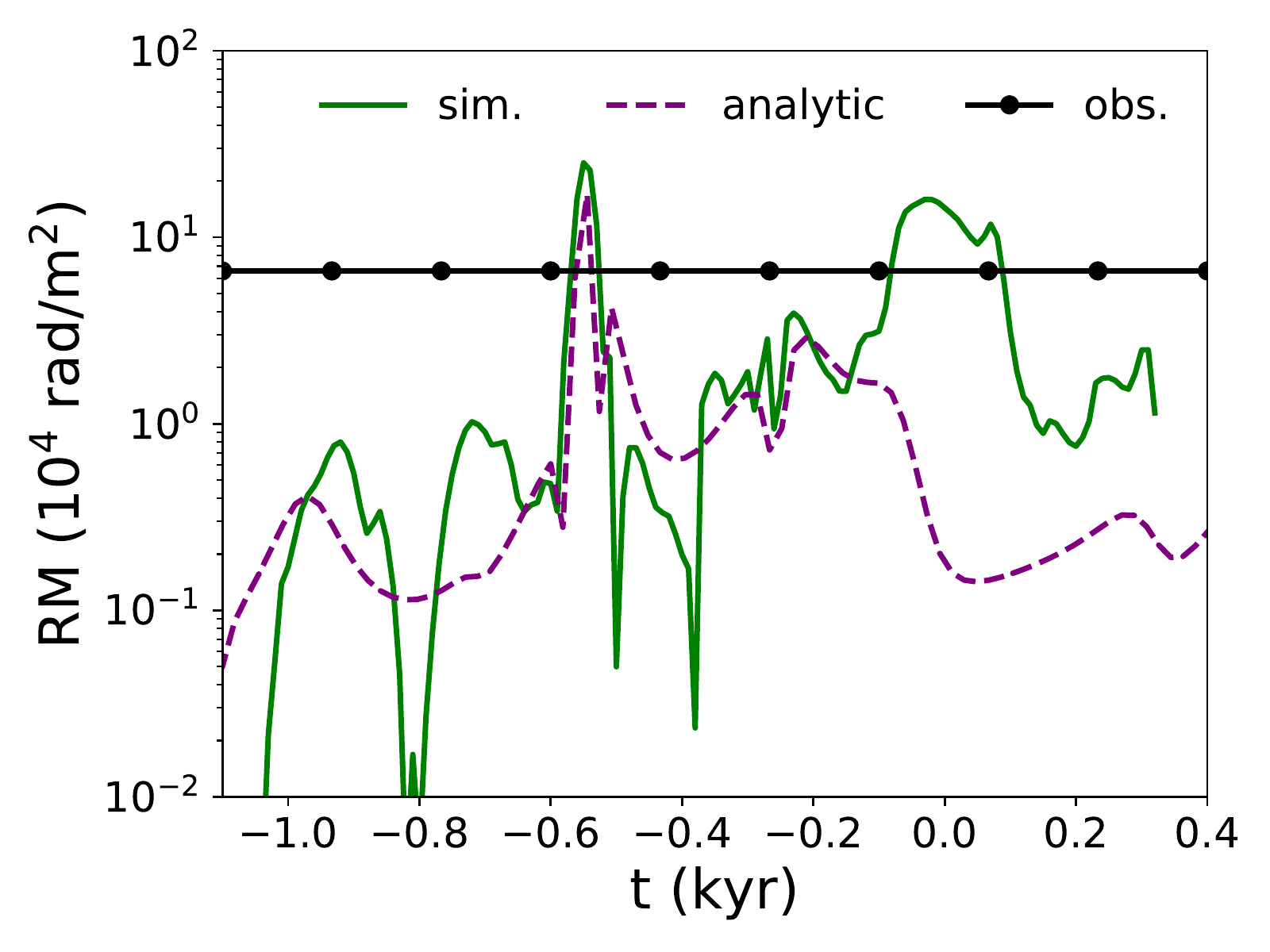}
\includegraphics[width=0.45\textwidth]{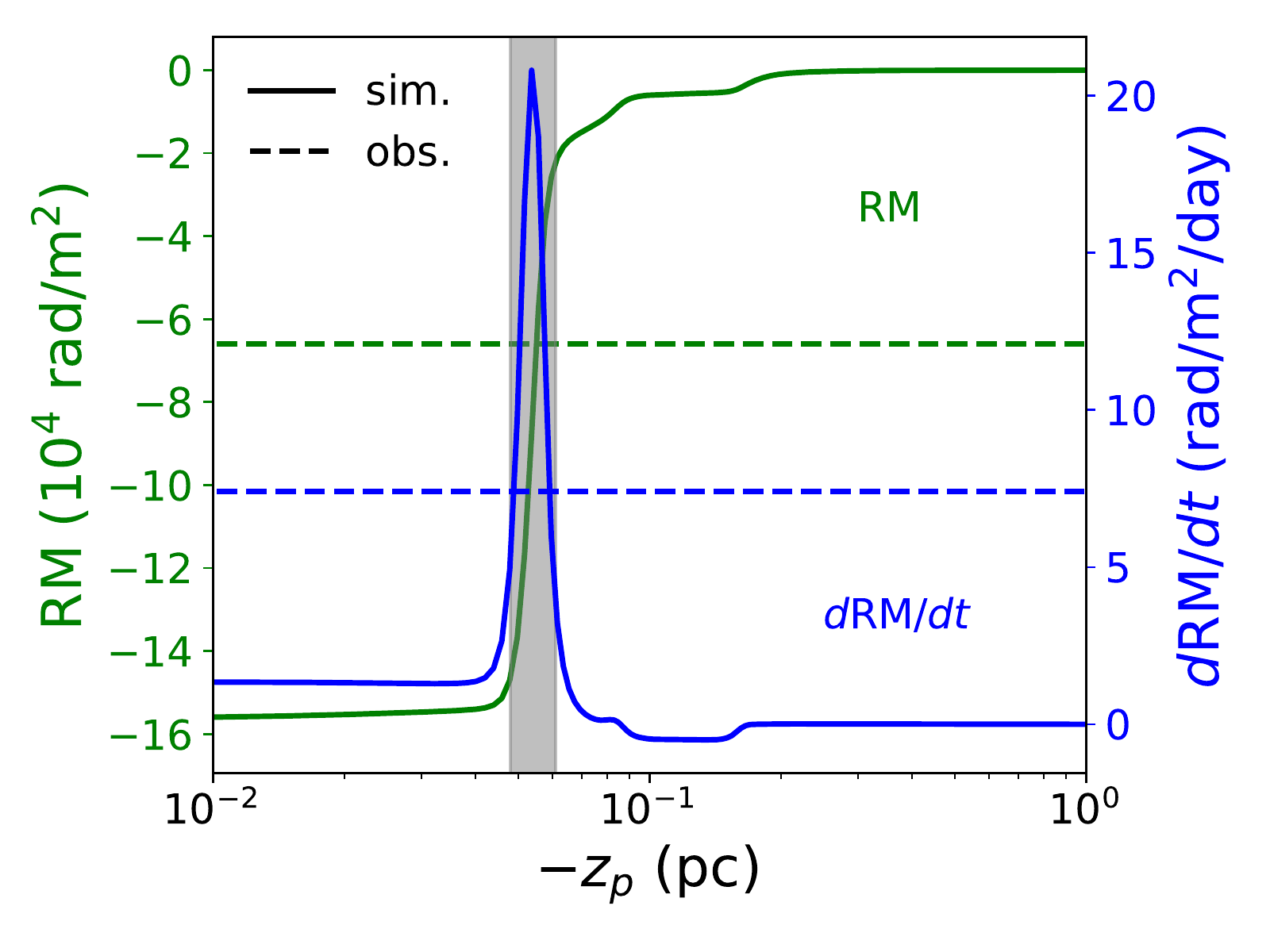}
\caption{Top: Absolute value of the RM as a function of time at the pulsar's present day LOS calculated from our fiducial simulation (\S \ref{sec:MHD}) compared to our analytic, isolated stellar wind model (\S \ref{sec:analytic}).  Also plotted is the present day magnitude of the pulsar's RM, $\approx 6.6 \times 10^4$ rad/m$^{2}$.  Bottom: RM (solid green) and the time rate of change of the RM (solid blue) at $t=0$ as a function of the $z$-coordinate of the pulsar, $z_p$, compared the observed RM (dashed green) and time variability (dashed blue, \citealt{Desvignes2018}). The shaded gray area represents the region in between the two disc stars, E32 and E40, where the two winds are shocking and enhancing the RM.   }
\label{fig:RM_time}
\end{figure}

The top panel of Figure \ref{fig:RM_time} shows the RM as a function of time at the $t=0$ location of the pulsar calculated from our fiducial simulation 
as $z_p$, the LOS position of the pulsar, $\rightarrow \infty$. Since neither the location nor velocity of the pulsar along the LOS is known, we fix its position to show how the RM varies at its current location. The bottom panel of Figure \ref{fig:RM_time} shows how $z_p$ affects the RM and its gradient in time.  In this fiducial model, we find that we can roughly match the observed value of the pulsar RM and its gradient, if it is located within or behind the stellar disc.  
Although our predicted dispersion measure (DM) for the pulsar is of order $\sim 50$ pc/cm$^{3}$, $\ll$ the observed value of $\sim$ 1700 pc/cm$^{3}$ (as expected for a DM dominated by a screen far from the pulsar), its gradient in time can be large enough ($\sim 2.5$ pc/cm$^{3}$/yr) to plausibly account for the $\sim 0.06\%$ change observed over a four year period \citep{Desvignes2018}.

\begin{figure}
\includegraphics[width=0.49\textwidth]{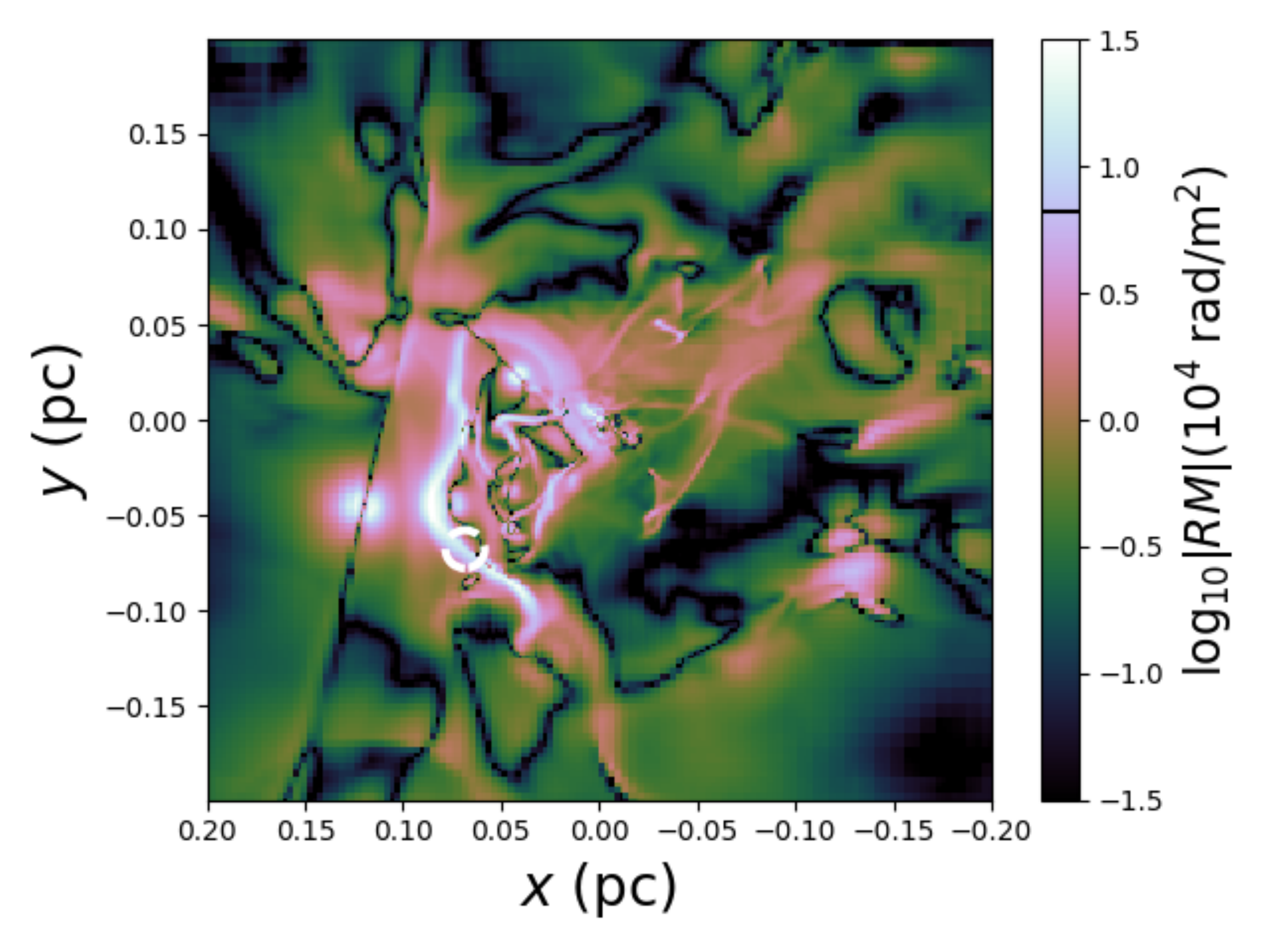}
\caption{Absolute value of the RM calculated from infinity as a function of location in the plane of the sky for the galactic centre.  The white circle represents the present day location of the magnetar, while the black line on the colorbar indicates the observed value of its RM. The origin is Sgr A*. Due to a shock between the winds of the stars E32 and E40 (aka 16SE1 and 16SE2), our fiducial model produces a RM large enough to explain the observed value at the pulsar's location.  Such a large RM, however, is not ubiquitous to the central $\sim 0.1$ pc but occurs in only in a small portion of the domain, requiring a fortuitous alignment of the pulsar with either a wind or a shock.  The probability of this occurring could be enhanced if the pulsar is on an orbit within or nearly within the clockwise stellar disc. Other pulsars detected in the future would likely have lower RMs. }
\label{fig:RM_map}
\end{figure}

Also shown in the top panel of Figure \ref{fig:RM_time} is the RM for the pulsar calculated from the analytic isolated wind model as $z_p \rightarrow \infty$ (Equation \ref{eq:RM_cgs} summed over all the stars), which neglects the effects of wind-wind interaction. Certain peaks in the RM (e.g., those at $\sim$ -0.5 kyr and $\sim$ -0.2 kyr) are well reproduced by the analytic model while others are seen only in the simulation (e.g., those at $\sim$ -0.7 kyr and the present day). The latter are caused by strong shocks between nearby winds, while the former are caused by winds located far from other stars.  At the present day the RM is dominated by a radiative, magnetic pressure dominated shock between the winds of E32 and E40, with a post-shock region characterized by $|B_z| \approx 10$ mG, $n_e \approx 3000$ cm$^{-3}$, $T \approx$ 10$^4$ K, and a LOS width of 0.01 pc. This shock is clearly seen in the 2D map of RM (Figure \ref{fig:RM_map}). We note that radiative cooling is not required for a large RM, which we have confirmed with a simulation that neglects cooling yet still has a comparable RM at the pulsar's LOS and across the domain.  

Both Figure \ref{fig:RM_time} and \ref{fig:RM_map} show that the large RM at the $t=0$ LOS of the pulsar is somewhat rare in both space and time. The typical value is closer to 
$\sim 1 \times 10^4$ rad/m$^{2}$.
This is true for all of the variants we simulated (fixing $\beta_w \sim 10-100$; for $\beta_w\gg 100$ we found no simulation with a large enough RM) and suggests that the current high value of the observed RM is the result of a chance alignment of the pulsar with the region associated with three disc stars in close proximity. This result is not strongly affected by the $\sim$ 30-60$\%$ uncertainties on the $t=0$ $z$-coordinates of these stars since the separations between them are predominantly perpendicular to the LOS.
\citet{Pulsar_loc} found that the proper motion of the magnetar is consistent with an orbit in the clockwise stellar disc.  Such an orbit might put the pulsar into more frequent alignment with closely interacting stellar winds and enhance the likelihood of observing a RM with the observed magnitude.  

\subsection{Rotation Measure of Sgr A*}

\begin{figure}
\includegraphics[width=0.45\textwidth]{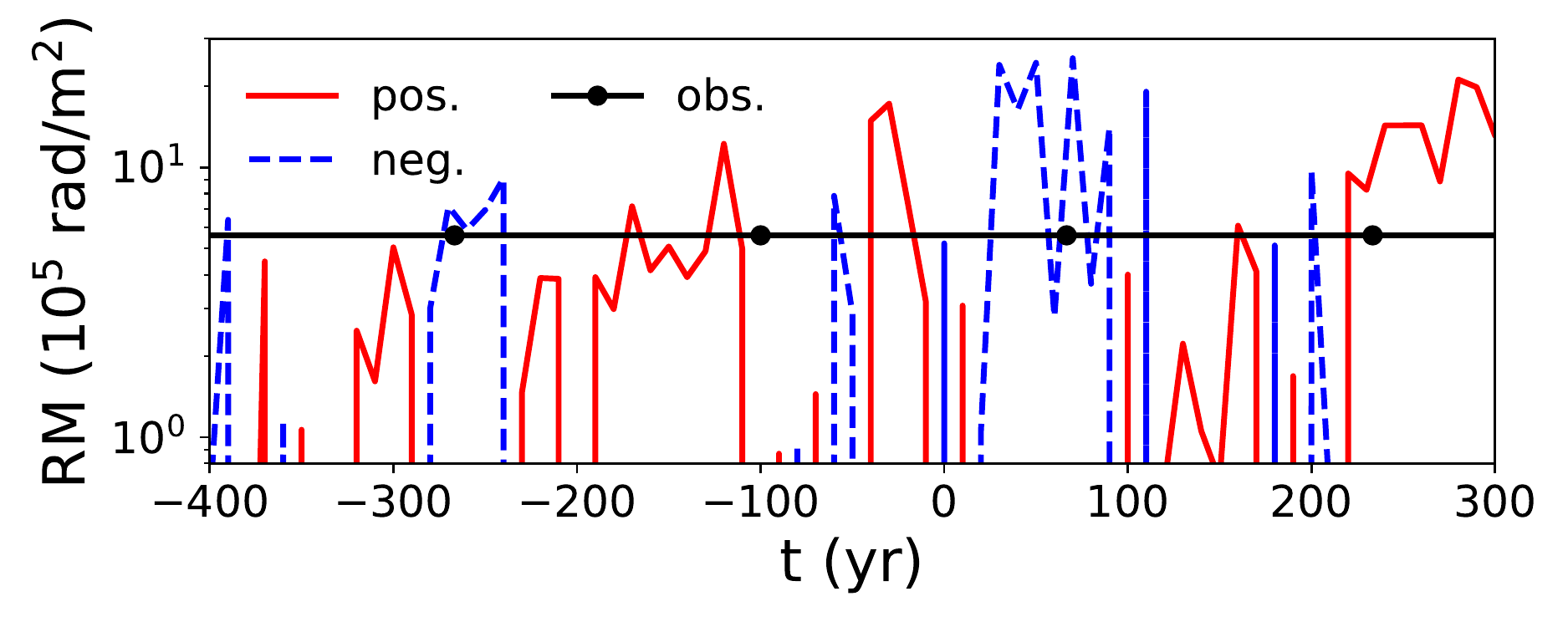}
\includegraphics[width=0.45\textwidth]{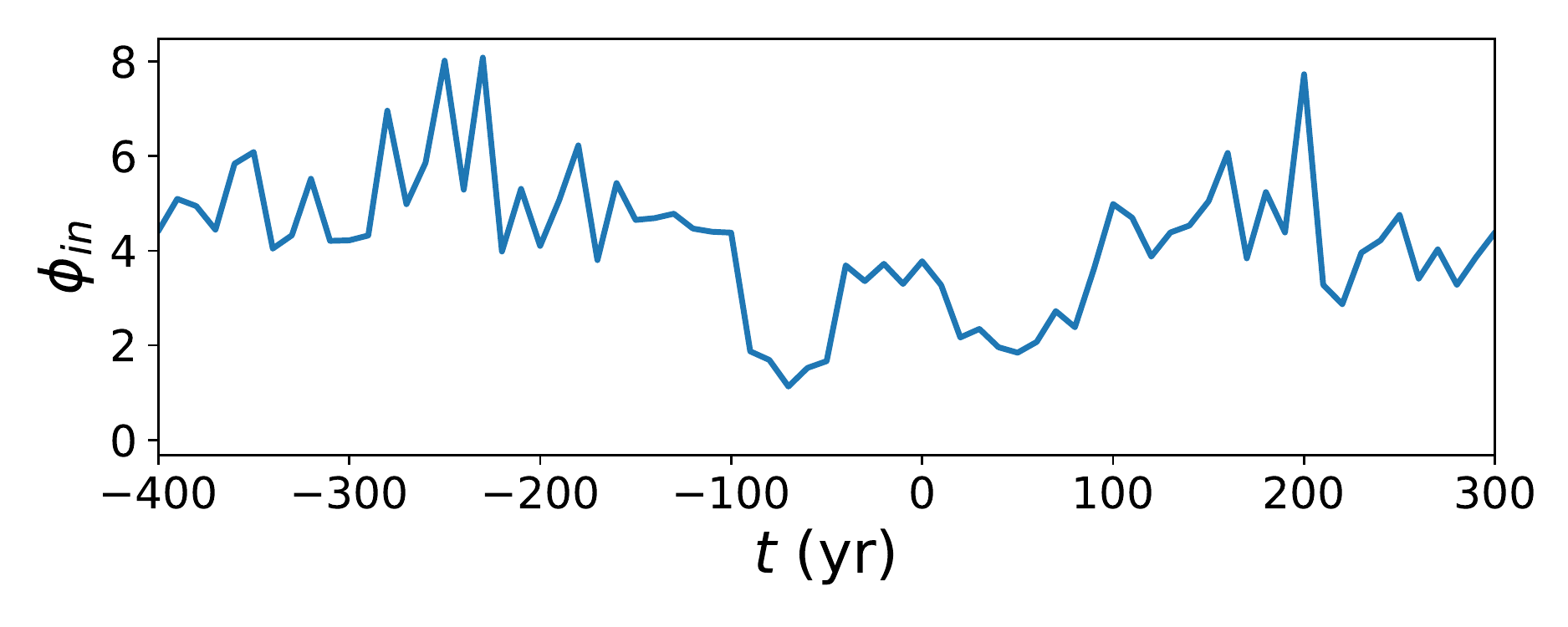}
\caption{Top: RM as a function of time for Sgr A* in our fiducial 3D MHD simulation. 
Solid red lines are positive RMs, while blue dashed lines are negative RMs.  Also plotted is the observed $t=0$ magnitude of the RM, $\approx 5.6 \times 10^5$ rad/m$^{2}$.  
The RM of our simulation is typically of order the observed value and can remain the same sign for intervals as short as a few years or as long as $\sim$ 100 years.  Though the RM and its smaller amplitude variability are set by the innermost region of the simulation ($r \sim 10^{-4}$ pc where the dynamical time is $\sim 3.5$ days), the timescale for the RM to change sign is set by the dynamical time at much larger radii.  Bottom: Dimensionless flux threading the inner boundary, $\phi_{in}$, as a function of time (Equation \ref{eq:phi_in}). This ordered field leads to a strongly magnetized accretion flow, $\phi_{in}\sim 2-6$, where $\phi_{in}\sim 50$ roughly corresponds to the MAD limit.  }
\label{fig:RM_sgra_time}
\end{figure}

The top panel of Figure \ref{fig:RM_sgra_time} shows the RM of Sgr A* as a function of time over a 700 year period.   In contrast to the pulsar's RM, which at the present day can be directly traced to the parameters of only $\sim 3$ of the winds, the precise behavior of the RM of Sgr A* near $t=0$ is a complicated function of the spin axes, mass loss rates, and wind speeds of all 30 of the stellar winds.  This is due to the fact that the RM of Sgr A* is set by the accreting material at the innermost region of our simulations at $r_{in} \approx 1.2 \times 10^{-4}$ pc, potentially a combination of gas from multiple winds.
Because of this, we focus our analysis on the general statistical behavior of the RM of Sgr A* instead of its specific behavior at $t=0$. 

The RM in Figure \ref{fig:RM_sgra_time} is a reasonable estimate even though our inner boundary does not extend all the way to the horizon. We expect the largest contribution to the RM to be set by the radius at which the electrons become relativistically hot, which is only slightly inside the inner boundary of our simulation.  
For the non-relativistic RM, $d(\textrm{RM})/d\log(r) \propto r n_e B_\parallel \propto r^{-1}$, where we have used the radial dependencies of $n_e \propto r^{-1}$ and $|\mathbf{B}| \propto r^{-1}$ observed in our simulation.  We have confirmed that this radial dependence of the RM is valid by running simulations with larger inner boundary radius.  Once $k_b T_e \sim m_e c^2$, however, the RM becomes suppressed by factors of $\Theta_e^{-2}$, where $\Theta_e \equiv k_b T_e/(m_e c^2)$.   At the innermost boundary of our simulation, $\langle \Theta_e \rangle \approx 0.9$, and thus we would expect the RM to be set 
by the plasma properties at $r \sim 10^{-4}$ pc (i.e. $\approx$1.2 $r_{in}$). The magnitude of this RM is comparable for simulations with $\beta_w=10$ and those with $\beta_w=1000$, and is thus only weakly dependent on the magnetization of the stars. 

A striking feature of the RM towards Sgr A* shown in Figure \ref{fig:RM_sgra_time} is the timescale for sign changes, ranging from $\sim$ 3-100 years, much longer than the timescale of a few days for small amplitude fluctuations.  Since the RM is dominated by scales $\sim$ $r_{in} \approx 1.2 \times 10^{-4}$ pc, this means that the magnetic field is coherent in sign over $\sim$ $500-10,000$ Keplerian orbits and thus this sign is set by the dynamics at larger radii.  We have confirmed this hypothesis by running simulations with different values of $r_{in}$, finding that although the magnitude of the RM scales as $r_{in}^{-1}$, the timescale for it to flip sign is roughly independent of $r_{in}$. Furthermore, this $\sim$ 3-100 year time scale is a robust result in simulations with a range of different wind parameters. We conclude that this is a generic prediction of our model due to the fact that the magnetic field is sourced by stellar winds at large radii.  We expect that continual monitoring of the RM of Sgr A* over $\sim$ 10s of years would reveal a similar level of variability as seen in Figure \ref{fig:RM_sgra_time} and, eventually, a sign change. Our simulations can plausibly explain the factor of $\sim$ 2 variability seen by ALMA over $\sim$ months between epochs in \citeauthor{Bower2018} (\citeyear{Bower2018}; results that appeared as this work was in press) though not the much more rapid variability seen over $\sim$ hours within epochs. No sign reversal was yet seen.

Finally, we note that our simulations display a highly ordered magnetic field in the inner accretion flow with $|\langle \mathbf{B}\rangle|/\sqrt{\langle \mathbf{B}^2 \rangle} \approx 0.3-0.4$, where $\langle\rangle$ denotes an average over all angles and radius for $r\lesssim0.03$ pc. As shown in the bottom panel of Figure \ref{fig:RM_sgra_time}, this ordered field corresponds to a net magnetic flux threading one hemisphere of the inner boundary of 
\begin{equation}
\phi_{in} \equiv \left. \frac{1/2\int |B_r| r^2 d\Omega }{ r\sqrt{|\dot M | v_{kep}}}\right|_{r = r_{in}} \approx 2-6,
\label{eq:phi_in}
\end{equation}
where $v_{kep}$ is the Keplerian velocity. We have found that this value is roughly independent of $\beta_w$, the orientation of the spin axes of the stars, and even $r_{in}$, so by extrapolating our simulations we robustly estimate $\phi_{in} \sim 2-6$ at the horizon.  Considering that a magnetically arrested state (MAD) of accretion begins when  $\phi_{in} \approx 50$ \citep{Sasha2011}, this is a fairly significant amount of magnetic flux that could result in the formation of strong jets. As defined, $\phi_{in}$ is positive definite.  The field responsible for this flux in the innermost regions of our simulation, however, reverses direction on roughly the same timescale as the RM in the top panel of Figure \ref{fig:RM_sgra_time}.  

\section{DISCUSSION AND CONCLUSIONS}
\label{sec:conc}
We have shown that for a reasonable set of parameters, magnetized, 3D simulations of wind accretion onto Sgr A* reproduce the large RM observed towards the galactic centre magnetar, PSR J1745-2900, and can even account for its relatively large temporal gradient.  Additionally,
we find that for the same parameters the RM towards Sgr A* in our simulation is provided by an ordered magnetic field at $r \sim 10^{-4}$ pc ($\sim$ 250 Schwarzschild radii) and is roughly consistent with the observed value. Sgr A*'s RM retains its sign for $\sim 3-100$ year periods (depending on the exact simulation parameters), also consistent with polarization measurements. What's more, our predicted X-ray luminosity at scales of $2-10''$ from Sgr A* is consistent with \emph{Chandra} observations, suggesting that the properties of the hot gas are being modeled faithfully.  

In our models it is likely that the pulsar itself is within the clockwise stellar disc and that the RM is probing a shocked region between two WR stars in that disc. On the surface, such an explanation for the exceptionally large observed RM would seem like a fortuitous coincidence; regions with such large RMs are fairly rare in space and time at distances from the black hole comparable to that of the pulsar's LOS (Figures \ref{fig:RM_time} and \ref{fig:RM_map}).  On the other hand, if, as proposed by \citet{Pulsar_loc}, the pulsar itself is on a bound, clockwise orbit in or near the disc, then proximity to regions of enhanced density and magnetic field would be more common.  

Alternative models for the RM of the pulsar cannot be ruled out.  \citet{Sicheneder2017} show that a chance alignment of the pulsar's LOS with an HII region much closer to Earth can reasonably explain both the observed DM and RM if the region is magnetized.  \citet{YZ2015b} argue that the RM could easily be provided by warm, ionized gas in Sgr A* west.  Another possibility is that while the pulsar's RM is local to the galactic centre, it is not local to the inner parsec.  This is suggested by the fact that two other pulsars in the galactic centre, J1746−2849 and J1746−2856, located 10-100 pc away from Sgr A* in projected distance, have RMs that are also fairly large, $\sim$ $10^4$  rad/m$^{2}$ \citep{Schnitzeler2016}.

In summation, we have presented a single numerical model that simultaneously explains the observed diffuse X-ray luminosity towards the galactic centre, the value and variability of the RM of the galactic centre magnetar, and the magnitude and constancy in sign of the RM of Sgr A*.  Continual monitoring of the pulsar's motion and acceleration, follow-up observations of the RM of Sgr A*, improved constraints on the mass-loss rates and wind speeds of the stellar winds, and magnetic field strength estimates based on observations of Zeeman splitting of the absorption lines for the three winds closest to the pulsar (i.e. the winds of E23, E32, and E40) will be important for testing the validity of this picture.  Our simulations predict a 2D RM map of the inner 0.5 pc of the Galaxy that can be used to test whether the RM is indeed produced local to Sgr A* using future pulsar detections.   We also predict that Sgr A* is typically accreting significant magnetic flux (though below the MAD limit), enough to potentially power strong magnetic outflows.    

\section*{ACKNOWLEDGEMENTS}
We thank J. Dexter, C. Law, R. Genzel, F. Yusef-Zadeh, and D. Mu\~{n}oz for useful discussions, as well as all the members of the horizon collaboration, \href{http://horizon.astro.illinois.edu}{http://horizon.astro.illinois.edu}t.  We thank the referee for a positive report. This work was supported in part by NSF grants AST 13-33612, AST 1715054, AST-1715277, \emph{Chandra} theory grant TM7-18006X from the Smithsonian Institution, a Simons Investigator award from the Simons Foundation, and by the NSF
through an XSEDE computational time allocation TG-AST090038
on SDSC Comet.   SMR is supported in part by the NASA Earth and Space Science Fellowship.  This work was made possible by computing time granted by UCB on the Savio cluster.

\bibliographystyle{mn2efix}
\bibliography{RM}

\end{document}